\documentclass[aps,prl,reprint,showpacs]{revtex4-1}

\usepackage[dvipdfmx]{graphicx}
\usepackage{mathrsfs} 
\usepackage[mathscr]{euscript}
\usepackage{amsmath,amssymb}
\usepackage{bm}
\usepackage{color}

\begin{document}

\title{Lattice Unruh effect and world-line entanglement for the XXZ chain}

\author{ Kouichi Okunishi and Kouichi Seki}

\affiliation{Department of Physics, Niigata University, Niigata 950-2181, Japan}

\date{\today}

\begin{abstract}
For the XXZ chain, we discuss the relation between a lattice version of Unruh effect and the ground-state entanglement on the basis of the corner Hamiltonian.
We find that the lattice Unruh temperature is interpreted as $\beta_\lambda = 2\pi/a$ with an effective acceleration $a=\pi/\lambda$, where $\lambda$ denotes the anisotropy parameter of the XXZ chain.
Using quantum Monte Carlo simulation for the corner Hamiltonian at $\beta_\lambda$, we demonstrate that world lines of spins surrounding the entangle point provides an intuitive understanding the quantum entanglement.
We also propose an XXZ-chain analogue of the detector for the thermalized entanglement spectrum with use of the angular time evolution defined by the corner Hamiltonian.
\end{abstract}


\maketitle

The concept of entanglement in quantum many-body systems has been providing deep insights for interdisciplinary research fields of physics.
A most striking example is the holographic principle for the entanglement entropy, which suggests that there is a common theoretical background among quantum many-body systems, quantum field theories and quantum/classical gravity.\cite{Ryu-Takayanagi}
Moreover, the entanglement analysis has been an essential tool in designing tensor network simulation algorithms for quantum and classical spin systems\cite{DMRG,DMRG2, CTMRG1,CTMRG2,MERA1}, which have also certain connections to the quantum information and the holographic entanglement entropy\cite{Swingle}.

A most typical quantum many-body system for understanding the quantum entanglement is the XXZ chain, where the integrablility reveals various fundamental properties of the quantum entanglement; 
The reduced density matrix of half-infinite bipartitioning of the XXZ chain can be exactly constructed through the corner transfer matrix (CTM) for the corresponding 6-vertex model\cite{Baxterbook,BaxterCTM1}, which enables us to extract the exact entanglement entropy\cite{Calabrese-Cardy} and the  asymptotic form of the entanglement spectrum \cite{Peschel-Kaulke-Legeza,OHA,Calabrese-Lefevre}.
Also, conformal field theories for the CTM geometry clarifies well-known behaviors of the entanglement spectrum and entropy in the critical regime.\cite{Peschel-Troung,Cardy-Tonni, Holzhey, Calabrese-Cardy}
Moreover, the CTM also becomes an essential ingredient in recent developments of tensor network simulations\cite{TPVA,TPVA_q,Orus,Corboz}. 

In this letter,  we focus on another key property of the CTM to reveal its interesting connection to quantum field theories in non-inertial frames;
The corner Hamiltonian ---the generator of the CTM equivalent to the bipartition entanglement Hamiltonian---  works as a lattice Lorentz boost operator with respect to the rapidity parameterizing the 6-vertex model.\cite{Sogo-Wadati, Thacker}
This suggests that the ground-state entanglement of the XXZ chain can be interpreted as a lattice version of the Unruh effect, that is the nontrivial equivalence between the usual vacuum of a quantum fiend theory and the thermalized states observed by a constantly accelerating observer, where the spectrum of the Lorentz boost operator plays also a key role \cite{Unruh, Fulling, Rev_Mod_Unruh}.

Using a world-line(WL) type quantum Monte Carlo (QMC)\cite{Kawashima-Harada,Syljuasen-Sandvik} for the corner Hamiltonian, we discuss that the reduced density matrix for the ground-state of the XXZ chain with the Ising-like anisotropy can be illustrated as superposition of WLs of spins surrounding the entangle point at the lattice Unruh temperature.
We then find that the scale factor of the imaginary angular time defines an effective acceleration in the lattice Unruh effect.
We also demonstrate that the thermal average of physical quantities and thermal entropy for the corner Hamiltonian respectively reproduces the ground-state expectation values and the bipartition entanglement entropy of the XXZ chain.
In analogy with the Unruh effect, moreover, we propose a spin-chain analogue of the Unruh-DeWitt detector, which may capture the thermalized spectrum of the reduced density matrix for the ground state.

Let us start with writing the XXZ chain Hamiltonian in the Ising-like regime as
\begin{equation}
{\cal H} =  J_\lambda \sum_{n=-L+1}^{L} \left[S_n^x S_{n+1}^x + S_n^y S_{n+1}^y + \Delta S_n^z S_{n+1}^z \right ]
\label{XXZ_H}
\end{equation}
where $\bm{S}$ are $S=1/2$ spin matrices, $L$ is a positive integer representing the system length.
The exchange coupling and the anisotropy are respectively parameterized as 
$J_\lambda =\frac{2}{\sinh \lambda}$
and  
$\Delta =\cosh \lambda$ with $\lambda > 0 $.
Note that $ \lambda \to \infty$ ($+0$) corresponds to the Ising (Heisenberg) limit.
In addition, we basically assume the open boundary conditions at the edges of the chain.
In the following,  we assume $n \ge 1$($n \le 0$) as the system (reservoir) part, where $n=0$ corresponds to the entangle point.

For the bipartition entanglement of the XXZ chain, an important implication of the integrability is that the ground state of  Eq. (\ref{XXZ_H}) is equivalent to the maximum-eigenvalue eigenvector of the transfer matrix of the 6-vertex model, which can be directly constructed through CTMs in the bulk limit.(See supplementary material)
We can then write the reduced density matrix for the bipartition of the chain as
\begin{align}
\rho = \exp(- \beta_\lambda {\cal K})/Z
\label{rdm}
\end{align}
where 
\begin{equation}
{\cal K} \equiv  J_\lambda \sum_{n=1}^{L} n \left\{S_n^x S_{n+1}^x + S_n^y S_{n+1}^y + \Delta S_n^z S_{n+1}^z\right\},
\label{XXZ_K}
\end{equation}
is the corner Hamiltonian and $Z\equiv \mathrm{Tr} \exp(- \beta_\lambda {\cal K})$.
In addition, $\beta_\lambda \equiv 2\lambda$ denotes the effective temperature characterizing a strength of quantum effect.
Here, we also assume the open boundary conditions for ${\cal K}$.

For revealing the entanglement structure of Eq. (\ref{rdm}), an important feature of ${\cal K}$ is that it satisfies the commutation relation as the lattice Lorentz boost operator with respect to the rapidity(See supplementary material).
In analogy with the Unruh effect\cite{Fulling,Unruh}, we then interpret Eq. (\ref{rdm}) based on ${\cal K}$ as a lattice version of Unruh effect, where the effective temperature $\beta_\lambda$ defines a measure of ``distance" from the classical limit.
However,  analytic calculation of eigenvectors of Eq. (\ref{rdm}) is a difficult  problem, although the exact spectrum of the corner Hamiltonian in the bulk limit was obtained\cite{BaxterCTM1}.
In this sense, numerical investigation for Eq. (\ref{rdm}) is essential for intuitive understanding of its entanglement structure.
Here, note that the effect of the open boundary for ${\cal K}$ is irrelevant in the bulk limit, as far as the Ising like regime where the correlation length is finite is concerned.

\begin{figure}[bt]
\includegraphics[width=6.5cm]{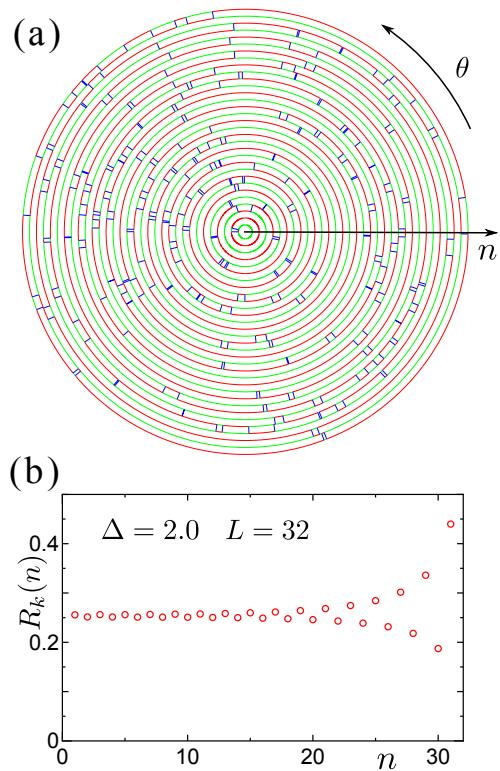}
\caption{(Color online) (a) A typical snapshot of WLs for the corner Hamiltonian at $\beta_\lambda$ with $\Delta=2$($\lambda=1.316\cdots)$ and $L=32$.
Red and green circles represent WLs carrying $S^z =\pm 1/2$. 
Short blue kinks connecting two adjacent circles indicate swapping of WLs due to the quantum fluctuation. 
(b) $R_{\rm k}(n)$ represents the normalized number of kinks between two adjacent circles of $n$ and $n+1$. 
$R_{\rm k}(n)$ becomes flat near the center of the snapshot of panel (a).}
\label{snapshot}
\end{figure}

Since the corner Hamiltonian ${\cal K}$ is an inhomogeneous Heisenberg model,   we can straightforwardly apply the loop-type algorithm of QMC to Eq. (\ref{rdm}), which enables us to directly generate typical WL configurations carrying $S^z = \pm 1/2$ at a finite temperature.(Supplementary material)
Then, a characteristic property of the corner Hamiltonian is that the scale of bond energy increases in proportion to the site index $n$, implying that the local imaginary time at $n$th site runs in $0 \le  n \tau \le   n \beta_\lambda $. 
We then introduce a normalized angle variable $\theta$ ($0\le \theta \le2\pi$) as $ \theta =a \tau $, where $a$ is a scale factor defined by
\begin{align}
a = \frac{2\pi}{\beta_\lambda} =\frac{\pi}{\lambda} \,.
\label{accela}
\end{align}
Note that the WL representation based on the $\theta$ variable is associated with an imaginary angular time of the angular quantization based on the Lorentz boost operator in the quantum field theory for the Rindler wedge\cite{Fulling}.
As will be discussed later, further, $a$ can be viewed as an effective acceleration constant in the lattice Unruh effect.

In Fig. \ref{snapshot}(a), we show a typical snapshot of WLs for $\Delta=2$ and $L=32$, where the red and green lines respectively represent WLs of $S^z = +1/2$ and $-1/2$, while the blue short lines in the radial direction indicate locations of kinks, i.e. swapping of the adjacent WLs due to the quantum fluctuation. 
Since the XXZ chain has the U(1) symmetry, the WLs  never terminate and thus always draw closed loops surrounding the entangling point($n=0$), reflecting the periodic boundary in the imaginary time direction.  
This implies that the fluctuating WLs winding around the entangle point literally represent the ground-state entanglement of the uniform XXZ chain.

In the Ising limit($\lambda \to \infty$),  WLs draw circles with no kink,  which indicate that there is no entanglement in the system.
In terms of the Unruh effect,  this classical limit corresponds to $a=0$, where an observer is classically separated from the reservoir part.
As $\Delta$ decreases,  number of kinks originating from the XY term increases, which induce nontrivial quantum entanglement in the system.
For the finite $\beta_\lambda$, the couplings in ${\cal K}$ increasing with respect to $n$ suggests that the center region of the system is in relatively high temperature,  while the outer region is in relatively low temperature.
If we write the number of kinks between the two adjacent circles of $n$ and $n+1$ as  $ N_k(n)$, then, $N_k(n)$ increases toward the outer region.
However,  the length of the WLs is proportional to the site index $n$.
An essential feature of the kink number at the Unruh temperature is that these two effects for kink density are nontrivially balanced.
In Fig. \ref{snapshot} (b), we show the normalized kink density $R_k(n) \equiv \langle N_k(n)\rangle / n$ for $L=32$ computed by QMC, which  actually demonstrates that  $R_k(n)$ is uniform for $n \lesssim 10$ where the boundary effect form the outer edge is negligible.
This implies that Eq. (\ref{rdm}) basically reproduces the uniform ground state of the Hamiltonian $\cal H$, except for the finite size effect from the outer edge.
Here we comment that if the temperature deviates from $\beta_\lambda$, the kink density exhibits anomalous behaviors in the vicinity of the center of the circles, reflecting a conical singularity at the center of the world sheet.

\begin{figure}[tb]
\includegraphics[width=7cm]{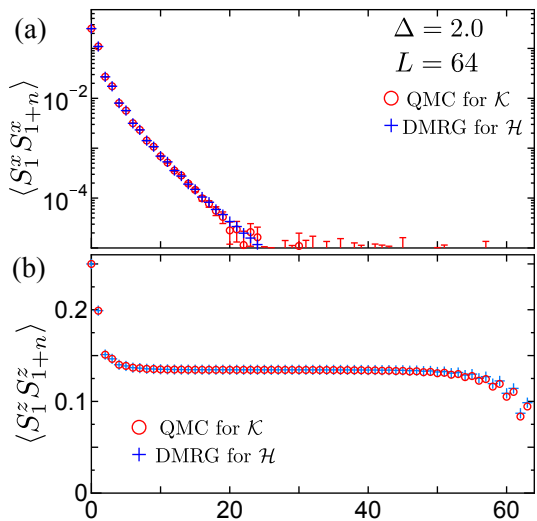} 
\caption{(Color online) Correlation functions for $\Delta=2$ and $L=64$: (a) $\langle S_1^x S_{1+n}^x \rangle$ and (b) $\langle S_1^z S_{1+n}^z \rangle$. 
Open circles indicate QMC results for ${\cal K}$ and cross symbols represent DMRG results for the ground state of ${\cal H}$.}
\label{figcor}
\end{figure}

We can also demonstrate the correspondence  of correlation functions between Eq. (\ref{rdm}) and the ground state of ${\cal H}$.
In Fig. \ref{figcor},  we show $\langle S_1^z S_{1+n}^z \rangle$ and  $\langle S_1^x S_{1+n}^x \rangle$ computed with QMC  for Eq. (\ref{rdm})  and those for the ground-state of $\cal H$ directly computed with density-matrix renormalization group (DMRG)\cite{DMRG}.
In the figure,  the QMC results for $\cal K$ are clearly consistent with the DMRG results for $\cal H$ within numerical accuracy.
Here, note that statistical errors in QMC results are of order of $10^{-5}$, which is basically negligible in the dominant scale of Fig. \ref{figcor}.

\begin{figure}[tb]
\includegraphics[width=7cm]{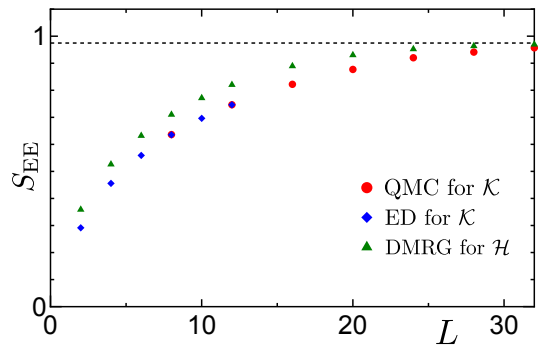} 
\caption{(Color online) Entanglement entropy for $\Delta=2$ up to $L=32$.
Open circles indicate QMC results for ${\cal K}$, where error bars are smaller than the symbol size. 
Exact diagonalization results for ${\cal K}$ up to $L=12$ are also shown as blue diamond symbols.
DMRG results of the entanglement entropy for the ground state of ${\cal H}$ are presented as green triangles for comparison.
The horizontal dotted line indicates the exact value $S_{\rm EE}=0.9747\cdots $.
}
\label{figEE}
\end{figure}

In addition to the observable quantities, it is also possible to extract the entanglement entropy from the WLs winding around the entangle point. 
Using Eq. (\ref{rdm}), we can straightforwardly obtain
\begin{align}
S_{\rm EE}= -{\rm Tr}_S[ \rho \log \rho]  
= \beta_\lambda \langle {\cal K} \rangle + \log Z 
\label{EE} 
\end{align}
which is nothing but the thermal entropy for the corner Hamiltonian ${\cal K}$.
Although precise evaluation of $S_{\rm EE}$ with QMC is a subtle problem for a large system size,  we perform numerical integration of the specific heat for $\cal K$ to estimate $S_{\rm EE}$ up to $L=32$. (See supplementary material).
Figure \ref{figEE} shows the entanglement entropy estimated with QMC for $\Delta = 2.0$.
We have confirmed that the QMC results are consistent with the exact-numerical-diagonalization results for $\cal K$ up to $L=12$, which are plotted as blue diamond symbols.
In Fig. \ref{figEE}, moreover, we also show the entanglement entropy directly computed with DMRG for the uniform Hamiltonian $\cal H$.
As $L$ increases, both of the QMC and DMRG results consistently converge toward the bulk value $S_{\rm EE}=0.9747\cdots$ extracted from the spectrum of the CTM\cite{BaxterCTM1, Peschel-Kaulke-Legeza, OHA}, although slight deviations due to the finite size effect remain up to $L=32$.

Here, we should comment on the above finite-size effect for $S_{\rm EE}$, which originates from the difference of the world-sheet geometries between the reduced density matrix of Eq. (\ref{rdm}) and the ground-state wavefunction for the uniform chain.
For the former case, the shape of the world-sheet of Fig. \ref{snapshot}(a) is basically a disk with the free boundary for the outer edge.
Whereas,  for the latter case, the ground-state wavefunction is represented by the half-infinite world sheet along the imaginary time direction.
For a relatively small $L$,  this geometric difference may emerge in the entanglement entropies as a finite-size effect as in Fig. \ref{figEE}.
If the system size is sufficiently large beyond the correlation length, the finite-size effect becomes negligible.
We have actually confirmed that the QMC and DMRG calculations converge to the exact bulk value for $\Delta=3.0$ within $L=20$,  where the correlation length is much shorter than $\Delta=2$.

We next discuss how to detect the thermalized spectrum of the reduced density matrix.
For the Unruh effect in the continuous space time, one can set up a constantly accelerating observer, which is described by the right-Rindler-wedge coordinate,
\begin{align}
x = r \cosh(a\eta)\, ,\quad  t = r\sinh(a \eta)\,, \label{rindler}
\end{align}
where $\eta$ denotes the proper time and $r$ is the spatial distance from the entangle point at $\eta =0$.
Note that $a$ in Eq. (\ref{rindler}) denotes the acceleration of the observer. 
We consider such an Unruh-DeWitt detector as harmonic oscillator coupled with a scalar field $\phi(x(\eta),t(\eta))$ along Eq. (\ref{rindler}).\cite{Unruh,DeWitt}
Then, the excitation rate of the accelerating detector is proportional to the $\eta$-integration of the Wightman function $\langle \phi(x(\eta),t(\eta)) \phi(r,0)\rangle$ along the trajectory (\ref{rindler}).\cite{Birrell,Brout,Rev_Mod_Unruh}
In this sense,  the correlation function with respect to the proper time involves essential information of the thermalized spectrum.

For the XXZ case, we could not define the literally accelerating observer.
However, we can exploit the fact that the scalar filed along Eq. (\ref{rindler}) can be formally written by the $\eta$-dependent Lorentz transformation,  $\phi(x(\eta),t(\eta)) = e^{-ia\eta K} \phi(r,0)e^{ia\eta K}$, where $K$ denotes the Lorentz boost operator for the scalar field, and $r\equiv x(0)$ is the distance from the entangling point at $t=0$.
The corner Hamiltonian is the lattice Lorentz-boost operator for the rapidity.
For the XXZ chain, thus, we may rather define a ``local spin" coupled with a detector as
\begin{align}
S_n^\mu(\eta) = e^{-ia \eta{\cal K}} S_n^{\mu} e^{ia\eta{\cal K}}\,,
\label{S_detector}
\end{align}
where $\mu\in x$ or $z$, $n$ corresponds to $r$ in Eq. (\ref{rindler}), and $\eta$ denotes the angular time scaled with the effective acceleration constant of Eq. (\ref{accela}).
Since $[S_n^\mu, {\cal K}]\ne 0$, the effective site range of $S_n^\mu(\eta)$ moves in the system, as $\eta$ increases.
Then, the detector coupled with $S_n^\mu(\eta)$ captures the autocorrelation function with respect to the angular-time evolution,
\begin{align}
G_n^\mu(\eta)
\equiv \frac{ {\rm Tr}\, S^{\mu}_n(\eta) S^{\mu}_n(0) e^{-\beta_\lambda {\cal  K}} }{Z} \, ,
\end{align}
which can be straightforwardly evaluated with Eq. (\ref{rdm}) in the basis diagonalizing ${\cal K}$.
However, we can also evaluate $G_n^\mu(\eta)$ for $n=1$ in the framework of DMRG for the ground state of the uniform XXZ chain, where we already have the eigenvalue spectrum of the reduced density matrix and the corresponding singular vectors.
Here, it should be noted that the singular vectors in DMRG play the same role as the Bogoliubov transformation  relating the field in the Minkowski space time to that in the Rindler coordinate.

\begin{figure}[tb]
\includegraphics[width=7cm]{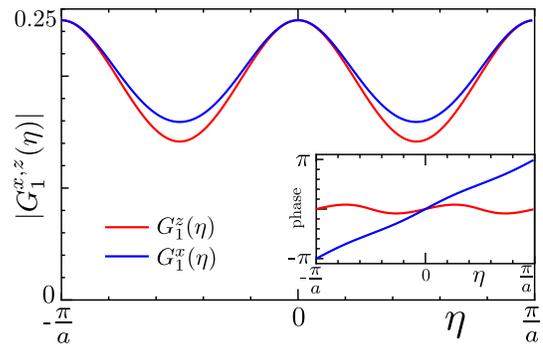} 
\caption{(Color online) Amplitude of the autocorrelation functions $G_1^{x,y}(\eta)$ for $\Delta=2.0$ and $L=64$.
Inset: phase of $G_1^{x, y}(\eta)$.}
\label{figauto}
\end{figure}

In Fig. \ref{figauto}, we show DMRG results of the autocorrelation function for $\Delta=2.0$ with $L=64$.
At $\eta=0$, we have the classical value $G_1^\mu(0)=1/4$, which corresponds to $a=0$. 
As $\eta$ increases, the amplitude and phase of $G_1^\mu (\eta)$ deviate from the classical value, implying that $G_1^\mu(\eta)$ actually captures the quantum entanglement.
We have confirmed that this deviation becomes more significant as $\lambda$ approaches to 0.
A particular behavior of $G_1^\mu(\eta)$ is that it exhibits $2\pi/a$ periodicity, which is distinct from the original Unruh effect for the continuous field.
This periodicity can be attributed to the lattice momentum $k$ of the spin waves in the XXZ chain, because the angular time $\eta$ is related to the rapidity $\alpha$ used in the coordinate Bethe ansatz through
$
e^{ik} = \sinh((\lambda+i \alpha)/2)/\sinh((\lambda-i\alpha)/2) \,
$
with $\alpha=a\eta$, where $-\pi \le \alpha\, ,  k < \pi$ (e.g. see Eq. (21b) in Ref. [\onlinecite{Yang-Yang}]).
Thus the nontrivial period for the $\eta$ evolution can be also read as a consequence of the rapidity modulation by the effective acceleration.  
Here, we note that, if $L$ is relatively  small compared with the correlation length of the system, the $2\pi/a$ periodicity in  Fig. \ref{figauto} is perturbed by the finite-size effect.

Finally, let us unify roles of $\tau$ and $\eta$ in terms of the 6 vertex model.
In the context of the corner Hamiltonian analysis, $\tau(=\theta/a)$ corresponds to  the imaginary angular time and $\eta$ describes the real angular time evolution.
We can relate these two variables with the rapidity $u$ through $u = a(\tau + i \eta) $, which parameterizes the Boltzmann weight of the 6-vertex model\cite{Baxterbook}.
For a given $\lambda( >0 )$ in the antiferroelectric regime (the Ising-like regime of the XXZ chain),  the range of $u$ where the Boltzmann weight is real positive is restricted in $0\le u \le \lambda$.
In this sense, the rapidity shift operator $e^{-u \cal K}$ for the row-to-row transfer matrix of the right half(right Rindler wedge) is physically relevant within  $0\le u \le \lambda$.
Taking account of the contribution from the left half of the row-to-row transfer matrix(left Rindler wedge),  the effective range of $u$ in  the reduced density matrix of Eq. (\ref{rdm}) turns out to be $2\lambda(=\beta_\lambda)$, which leads $0 \le \tau <2\pi/a$.
As mentioned in the previous paragraph, on the other hand, the imaginary part of $u$, i.e. real angular time $\eta$,  corresponds to the rapidity of the Bethe ansatz, where the $2\pi$ periodicity originating from the lattice momentum, implying $ -\pi/a   \le  \eta  <  \pi/a  $.
In the Unruh effect for the scalar field,  the angular time defined by the Lorentz boost operator is scaled by the acceleration in the real and imaginary directions.
In this analogy, we can interpret the nontrivial scale factor $a$ for the XXZ chain as the effective acceleration.
Here, we should note that the anisotropy parameter $\lambda$ in the XXZ chain controls both of the effective acceleration and the mass gap, whereas the acceleration and the mass term are independent in the Unruh effect for the scalar field.
In this sense, the lattice Unruh effect of the XXZ chain is a nontrivial consequence of the interaction effect.

To summarize, we developed the lattice Unruh effect for the XXZ chain on the basis of the corner Hamiltonian.
In particular, we demonstrated that the effective acceleration is associated with the anisotropy parameter $\lambda$ through Eq. (\ref{accela}) and the quantum entanglement is described by WLs winding the entangle point, where the imaginary-angular-time evolution defined by the lattice-Lorentz-boost operator plays a significant role.
We think that this result is a numerically exact example of the path integral representation of a tensor network\cite{Miyaji-Takayanagi-Watanabe} for the biparition entanglement.
Then, how we can relate the WLs for ${\cal K}$ with those of ${\cal H}$ is an interesting future problem.\cite{Seki} 
We have also proposed a spin system analogue of the Unruh-DeWitt detector, which captures the entanglement spectrum through the real angular time evolution.
Although experimental verification of the original Unruh effect in quantum field theories is usually very hard, the present results interestingly suggest that the lattice Unruh effect based on ${\cal K}$ could be simulated with realistic experiments of quantum spin systems or ultra-cold atoms.

In this letter, we have concentrated on the Ising-like regime of the XXZ chain, since the corner Hamiltonian spectrum has the stable bulk limit.
However, the formulation based on the corner Hamiltonian is also possible for the critical regime ($-1<\Delta\le 1$), in which the continuum limit of the free fermion model is included\cite{Itoyama-Thacker,Lukyanov}. 
Then, the Unruh effect in the critical regime of the XXZ chain is an interesting problem, in connection with conformal field theories in the CTM geometry \cite{Peschel-Troung, Cardy-Tonni, Cho-Ludwig-Ryu}.
Also, the angular quantization approach to quantum field theories\cite{Brazhnikov} may be another important view point for through understanding of the entanglement in quantum many-body systems.
We believe that the lattice Unruh effect stimulates further investigations of the quantum entanglement and its related physics, from both of theoretical and experimental viewpoints.

This work was supported by JSPS KAKENHI, Grant Number 17H02931. 
The authors thank T. Nakamura for valuable comments.

\bibliography{ctm_qmc}

\newpage
\vspace{5mm}

\begin{widetext}
\section{Supplementary material }


In this supplementary material, we briefly summarize the fundamental properties of the 6-vertex model and the algebraic structure of the corner transfer matrix(CTM).
We also present details of world-line (WL) Quantum Monte Carlo (QMC) simulations.

\subsection{6-vertex model and corner transfer matrix}

Following Ref.\onlinecite{Baxterbook}, we write the Boltzmann weight of the 6-vertex model as
\begin{align}
W(\mu, \nu | \mu', \nu')= \hspace{5mm}
{\setlength\unitlength{2mm}
\begin{picture}(4,4)(0,-0.3)
\put(2,-2.1){\line(0,1){4}}
\put(4.0,0.0){\line(-1,0){4}}
\put(2,-3.2){\makebox(0,0)[b]{\scriptsize \mbox{$\nu$}}}
\put(4.4,0){\makebox(0,0)[l]{\scriptsize \mbox{$\mu'$}}}
\put(2,2.5){\makebox(0,0)[b]{\scriptsize \mbox{$\nu'$}}}
\put(-0.4,0){\makebox(0,0)[r]{\scriptsize \mbox{$\mu$}}}
\end{picture}
}\\
\nonumber
\end{align}
where the indices $\mu,\nu, \mu',$ and $ \nu'$ takes $+$ or $-$,  respectively corresponding to $S^z=+1/2$ or $-1/2$ in the context of the XXZ chain. 
Then, the vertex weight is explicitly parameterized as 
\begin{align}
& W(+,+|+,+) = W(-,-|-,-)=1 \\
& W(+,-|-,+) = W(-,+|+,-)= \frac{\sinh(u)}{\sinh(\lambda -u)} \\ 
& W(+,-|+,-) = W(-,+|-,+)=\frac{\sinh(\lambda)}{\sinh(\lambda -u)}  
\label{6vbw}
\end{align}
where $u$ is the rapidity.
Note that $\Delta \equiv \cosh\lambda$ corresponds to the anisotropy of the XXZ chain.
The row-to-row transfer matrix is written as 
\begin{align}
T(u) = \sum_{\{\mu\}} \prod_n W_n(\mu_n,\nu_n|\mu_{n+1},\nu_{n+1}) \label{rtrTM}
\end{align}
where $n$ denotes the site index.
Although  the periodic boundary is basically assumed for Eq. (\ref{rtrTM}), the boundary condition is not relevant in the following argument about CTMs in the bulk limit, as far as $\Delta>0$ ($\lambda > 0$).
The integrability of the 6-vertex model ensures that the row-to-row transfer matrices of different rapidities satisfy $[ T(u), T(u') ] =0$.
According to the Baxter's formula, then, the Hamiltonian extracted from Eq. (\ref{rtrTM}) is
\begin{align}
\tilde{\cal H}& = -\left.\frac{d}{du} \log T(u)\right|_{u=0} \nonumber  \\
 &=   - J_\lambda \sum_{n} \left[S_n^x S_{n+1}^x + S_n^y S_{n+1}^y - \Delta S_n^z S_{n+1}^z \right ] + {\rm const}\, . 
\label{XXZ_alt}
\end{align}
Using the local unitary, ${\cal U} = \prod_{n={\rm even}} e^{i\pi S^z_n}$, we can  invert the sign of the XY term in Eq. (\ref{XXZ_alt}) to obtain the XXZ Hamiltonian of Eq. (\ref{XXZ_H}) in the main text. 
Here, note $[ T(u), {\cal H}] =0$.

\begin{figure}[tb]
\includegraphics[width=5cm]{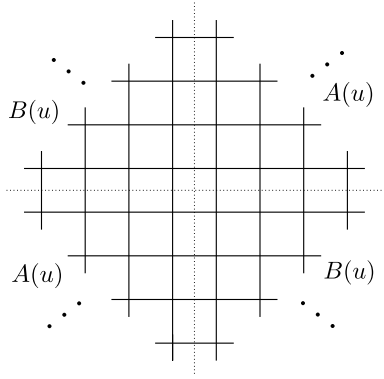} 
\caption{Graphical representation of CTMs. }
\label{figctm}
\end{figure}

The ground state of the XXZ model, i.e. the eigenvector of the maximum eigenvalue of the row-to-row transfer matrix can be constructed as
$|\Psi \rangle \sim \lim_{N \to \infty}  T^N(u) |\Psi_0\rangle $, where $ |\Psi_0\rangle$ is a certain initial vector that does not orthogonal to $ |\Psi \rangle$. 
Thus, the graphical representation of $|\Psi \rangle$ corresponds to the half-infinite plane with the initial condition $ |\Psi_0\rangle$.
As in Fig. \ref{figctm}, the CTM is defined for a quadrant of the square lattice, where its matrix elements correspond to the partition functions of the quadrant with given low and column spin configurations. 
This implies that the ground state of the XXZ model can be represented as a product of two CTMs in the bulk limit, $|\Psi \rangle \sim  B(u)A(u)$.
On the basis of Yang-Baxter relation, one can generally write $ B(u)A(v) \sim X(u-v)$, where $X$ is a matrix whose matrix elements depend only on  the rapidity difference $u-v$. 
Furthermore, the crossing symmetry of Eq. (\ref{6vbw}) ensures that the $\pi/2$ rotating of $A(u) $ gives $B(u) = A(\lambda - u)$.
These functional relations lead to
\begin{align}
A(u) \sim e^{-u {\cal K}}
\end{align}
where ${\cal K}$ represents the corner Hamiltonian defined by Eq. (\ref{XXZ_K}) in the main text. 
Thus, one can finally obtain the reduced density matrix, 
\begin{align}
\rho = B(u) A(u) B(u) A(u) \propto e^{-2\lambda {\cal K}}\,,
\label{rdm_sm}
\end{align}
 which defines the lattice Unruh temperature $\beta_\lambda \equiv 2\lambda$ independent of $u$.
 This equation corresponds to  Eq. (\ref{rdm}) in the main text.

We next review the algebraic relation of ${\cal H}$ and ${\cal K}$.
As shown in Refs \cite{Sogo-Wadati,Thacker},  the CTM $A(u)$ plays a role of the rapidity shift operator for the row-to-row transfer matrix,
\begin{align}
A(-v)T(u)A(v) =T(u+v)
\end{align}
or, equivalently 
\begin{align}
[{\cal K}, T(u)] = \frac{\partial}{\partial u} T(u)
\label{K_T}
\end{align}

Expanding the logarithm of the row-to-row transfer matrix with respect to $u$, 
\begin{align}
 \log T(u) = \sum \frac{I_n}{n!}u^n,
\end{align} 
we define a series of conserved quantities $\{I_n \}$ with $[I_n, I_m]=0$ for the 6-vertex model/XXZ chain.
In particular, $I_0$ and $I_1$ are respectively related to $I_0 = i P $(lattice momentum operator) and $I_1= - {\cal H}$ (the XXZ Hamiltonian).
From Eq. (\ref{K_T}), it follows that ${\cal K}$ plays a role of the ladder operator for $I_n$, 
\begin{align}
[{\cal K}, I_n] =  I_{n+1} \, .
\label{comK_In}
\end{align}
In particular, Eq. (\ref{comK_In}) includes 
\begin{align}
[P, {\cal H}] =0\,,\quad [ {\cal K,} P] = iH\, , \quad [{\cal K}, H] =  i \tilde{I}_2  \, ,
\end{align}
where $\tilde{I}_2= i I_2 = i \sum_{n=1} [h_{n,n+1}, h_{n+1,n+2}]$ with $h_{n,n+1}\equiv J_\lambda(S_n^x S_{n+1}^x + S_n^y S_{n+1}^y + \Delta S_n^z S_{n+1}^z)$.
The above commutation relation can be viewed as a lattice version of the Poincare algebra in 1+1 dimension\cite{Thacker}.
In particular, the corner Hamiltonian ${\cal K}$ corresponds to the lattice Lorentz boost operator.
Note that $P$, ${\cal H}$, $\tilde{I}_2$ are Hermitian.

\subsection{QMC details}

Assuming the $S^z$ basis representation, we briefly review the relation between matrix elements of the Ising-like XXZ chain and WL structures generated by QMC,\cite{Kawashima-Harada,Syljuasen-Sandvik} which is essential for discussing about Fig. \ref{snapshot} in the main text.
In this section, we write the local Hamiltonian of the XXZ chain as 
\begin{align}
{H}_{n,n+1} = J_{xy}(n)(S^x_{n}S^x_{n+1} + S^x_{n}S^x_{n+1})+J_{z}(n)S^z_{n}S^z_{n+1}
\label{QMC_H}
\end{align}
where we assume $0< J_{xy}(n) < J_z(n)$.
In WL QMC,  the partition function of the chain is represented as a trace of weighted WLs of spins with the continuous imaginary time index, on the basis of the path integral representation.  
For updating a WL snapshot with the loop-type algorithm, we consider two types of graph elements for WLs:
 ``binding graph'' and ``horizontal graph'' (See Fig. \ref{fig0}), which are respectively associated with  the $S^zS^z$ interaction term and the  XY term.
For a given WL configuration [Fig. \ref{fig0_1} (a)], the binding graph is placed for two adjacent WLs carrying the opposite spins (antiferromaginetic case) with the density $ (J_{z}(n)-J_{xy}(n))/2$ per unit length in the imaginary time direction.
Note that the total length of the imaginary time is given by the inverse temperature $\beta$.
Also, the horizontal graph is assigned for adjacent WLs having anti-parallel spin configurations with the density $J_{xy}(n)/2$, in addition to the kink place already included in the given WLs. [Fig. \ref{fig0_1} (b)]
After the allocation of the graph elements, we perform cluster analysis for WLs connected by the binding graphs. [Fig. \ref{fig0_1} (c)]
We then randomly update spin direction of each WL cluster [Fig. \ref{fig0_1} (d)], which gives a new configuration of WLs.
In the WL configuration of the XXZ chain, the total number of WLs carrying $S^z = \pm 1/2$ is conserved according to the total $S^z$ conservation.

\begin{figure}[tb]
\centering\includegraphics[width=7cm]{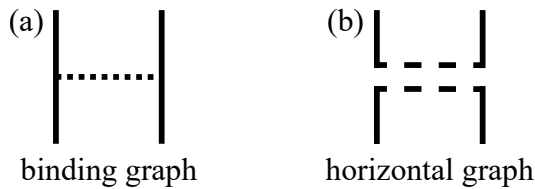}
\caption{
Graph elements for the Ising-like XXZ model.
(a) ``binding graph" anti-parallelly connects two adjacent WLs.
(b) ``horizontal graph" is inserted in two WLs carrying the anti-parallel spins. 
}\label{fig0}
\end{figure}

\begin{figure}[bt]
\centering\includegraphics[width=8cm]{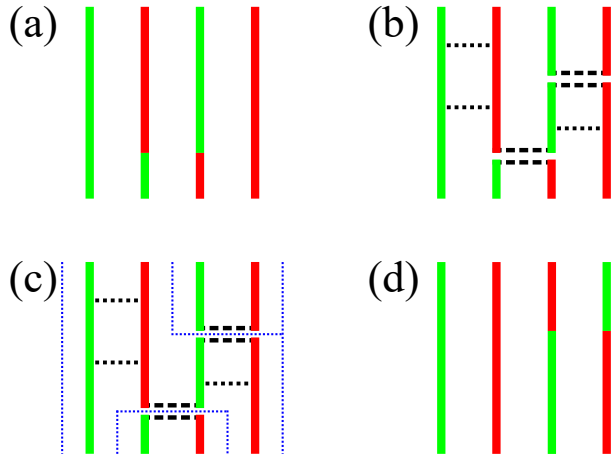}
\caption{
(Color online)
Update scheme of a WL configuration in the loop algorithm.
Red and green lines respectively represent WLs of up and down spins.
(a) A given WL configuration of the Ising-like XXZ model.
(b) Graphs assigned for the WL configuration.
(c) Cluster analysis for WLs connected by bind graphs.
The WLs surrounded by blue dotted lines form WL clusters.
(d) Up or down spins are randomly assigned for the clusters.
}\label{fig0_1}
\end{figure}

Using the loop type algorithm of the WL QMC for Eq. (\ref{rdm_sm}), we compute expectation values of various quantities.
In particular, ${\cal K} = \sum_n n h_{n,n+1}$ with $ h_{n,n+1}\equiv J_\lambda(S_n^x S_{n+1}^x + S_n^y S_{n+1}^y + \Delta S_n^z S_{n+1}^z)$, which leads $J_{xy}(n) \to n J_\lambda$ and $J_z(n) \to n J_\lambda \Delta$ in Eq. (\ref{QMC_H}).  
A typical sample number for average calculations is $1.0 \times 10^5$.
In Fig \ref{kink_bond}, we show results of the kink density $R_{\rm k}(n)$ and the normalized bond energy
 $\langle h_{n,n+1}\rangle$ for $\Delta = 2.0$.
Note that the number of kinks is closely related to the XY component of the bond energy.
Thus, $R_{\rm k}(n)$ basically shows very similar behaviors as $\langle h_{n,n+1} \rangle$ in Fig. \ref{kink_bond}.
In Fig. \ref{kink_bond}(a) and (c), size dependences of  $R_{\rm k}(n)$ and $\langle h_{n,n+1} \rangle$ at $\beta_\lambda$ are respectively shown.
As $L$ increases, the flat regions in the small $n$ side extend, implying that the uniform ground state of the original XXZ chain can be successfully reproduced.
We have actually confirmed that $\langle  h_{n,n+1} \rangle$ in the flat region of Fig. \ref{kink_bond} is consistent with the ground-state expectation value of the XXZ chain.

 Figures \ref{kink_bond}(b) and (d)  respectively show the average kink density $R_{\rm k}(n)$ and the normalized bond energy $\langle  h_{n,n+1}\rangle$ evaluated at temperatures different from $\beta_\lambda$.
If $\beta$ deviates from the lattice Unruh temperature, anomalous behaviors emerges toward $n=1$, implying that the uniformity near $n=1$ is broken down. 
As mentioned in the main text, the world sheet of $\beta$ deviating from $\beta_\lambda$ may have a conical singularity in the vicinity of the entangle point.
We think that the anomalous behaviors in Fig. \ref{kink_bond}(b) and (d) capture the conical singularity as the boundary effects.

\begin{figure}[tb]
\centering\includegraphics[width=14cm]{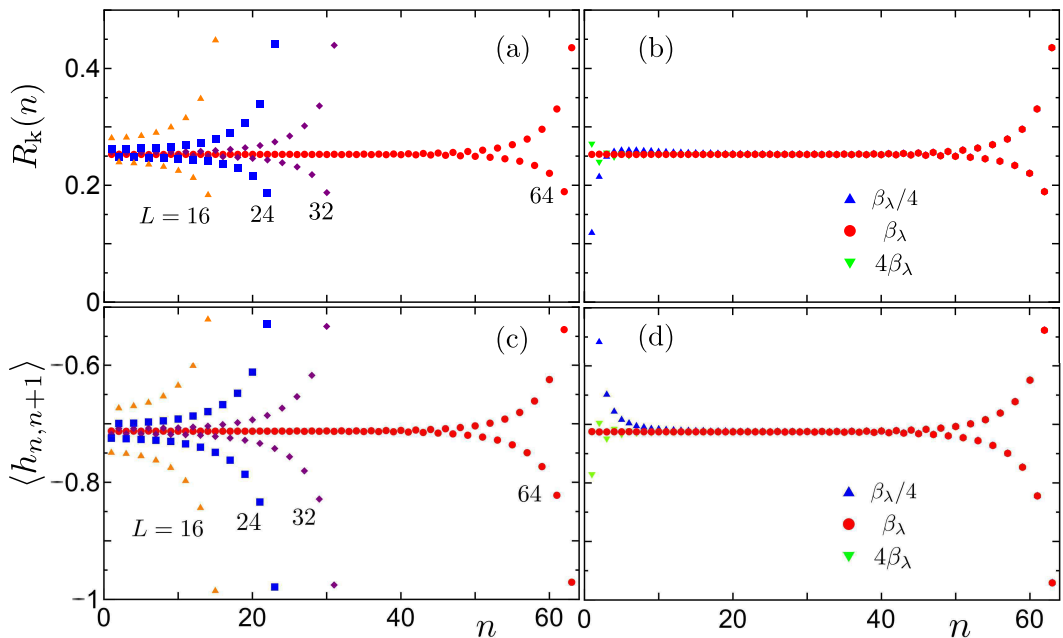}
\caption{
(Color online)
QMC results for ${\cal K}$ with $\Delta=2.0$.
(a) Size dependence of kink density $R_{\rm k}(n)$.
(b) Temperature dependence of $R_{\rm k}(n)$.
(c) Size dependence of the normalized bond energy $\langle  h_{n,n+1} \rangle$.
(d) Temperature dependence of $\langle  h_{n,n+1} \rangle$.
Error bares are smaller than the size of symbols in the figures.
}\label{kink_bond}
\end{figure}

\subsection{Estimation of entanglement entropy}

We evaluate the entanglement entropy $S_{\rm EE}$ of the XXZ chain as a thermal entropy for the corner Hamiltonian ${\cal K}$.
In general, a QMC calculation of the thermal entropy is also a subtle problem, because it is not an expectation value of a local quantity.
Here, we employ numerical integration of the specific heat $C_{\rm v}$ to evaluate $S_{\rm EE}$.
\begin{align}
S_{\rm EE} &= L\log2 -  \int_{T_\lambda}^{\infty}  \frac{C_{\rm v}}{T} dT = L\log 2 - \int_{\log T_\lambda}^\infty C_{\rm v} dx \label{CtoS1}  \\
& = L\log 2 - \left(\left[   E e^{-x} \right]^{\infty}_{\log(T_\lambda)} +  \int_{\log T_\lambda}^\infty E e^{-x} dx \right)
\label{CtoS2}
\end{align}
where $x \equiv  \log T$ and $L$ denotes the number of spins.
Note that, in the second line, $C_{\rm v}$ is converted to the internal energy $E$ with use of integration by part.

We perform WL QMC simulations for $\rho = \exp(-\beta {\cal K})/Z$ with the temperature being a free parameter. 
In Fig. \ref{Cv}, we show the internal energy per spin $E/L$ and the specific heat per spin $C_{\rm v}/L$ for $\Delta=2.0$ with  $L=8, 12, 16, 20, 24, 28$, and  $32$.
We also plot the lattice Unruh temperature $T_\lambda=1/\beta_\lambda= 1/2\lambda \simeq 0.3796 \cdots$ as a vertical dotted line. 
As $L$ increases, the amplitude of $E/L$ in the low temperature region increases.
At the same time, the peak temperature of $C_{\rm v}$ also shift to the high temperature side.
This is because the biggest scale of bond energies in ${\cal K}$ is proportional to $L$, which is distinct from the usual uniform spin chains.
Thus, the effective temperature scale of $T_\lambda$ becomes relatively low, as $L$ increases.
In Fig. \ref{Cv}(b), we see that data of $C_{\rm v}/L$ for the larger $L$ widely scatter in the low-temperature region.
Although the typical number of MC samples, $5\times 10^5$, was sufficient for precise estimations of the internal energy and spin correlation functions, the specific heat $C_{\rm v}$ estimated with QMC contains large statistical errors in the low temperature region.
On the other hand,  the internal energy $E$ is well estimated within the present QMC simulations, where errors  are of order of 6 digits.
However, we should note that the subtraction between the surface term and the last term in Eq.  (\ref{CtoS2}) may lose its numerical accuracy of $S_{\rm EE}$.
Thus, we need a careful treatment in the numerical integrations of Eqs. (\ref{CtoS1}) and (\ref{CtoS2}).

\begin{figure}[tb]
\centering\includegraphics[width=13.5cm]{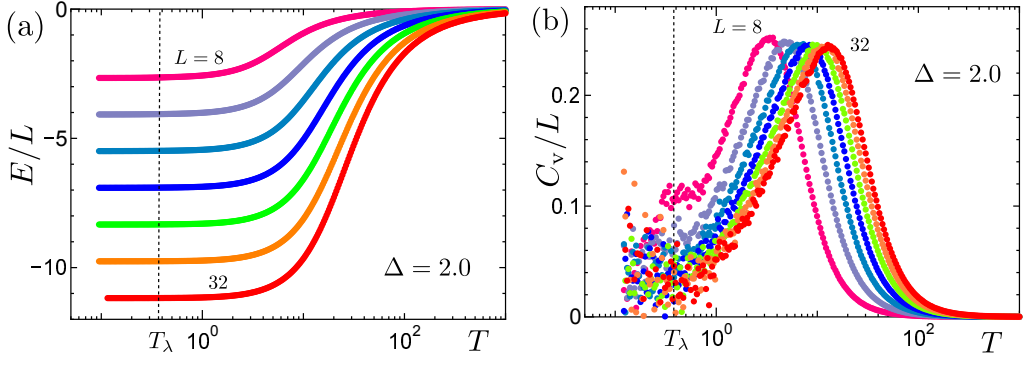}
\caption{
(Color online)
Internal energy per spin $E/L$ and specific heat per spin $C_{\rm v}/L$ for ${\cal K}$ with $\Delta = 2.0$.
(a) Temperature dependence of $E/L$.
The curves indicate $L=8$, 12, 16, 20, 24, 28 and 32 from top to bottom.
Error bares are negligible in the scale of the vertical axis.
(b) Temperature dependence of $C_{\rm v}/L$.
The curves indicate $L=8$, 12, 16, 20, 24, 28 and 32 from left to right.
Error bares are not shown for visibility of data.
A typical scale of the error bars in the low temperature regime is of order of the scattering data.
The vertical dotted line indicates the lattice Unruh temperature $T_\lambda=0.3796\cdots$ for  $\Delta = 2.0$.
}\label{Cv}
\end{figure}

\begin{figure}[tb]
\centering\includegraphics[width=14cm]{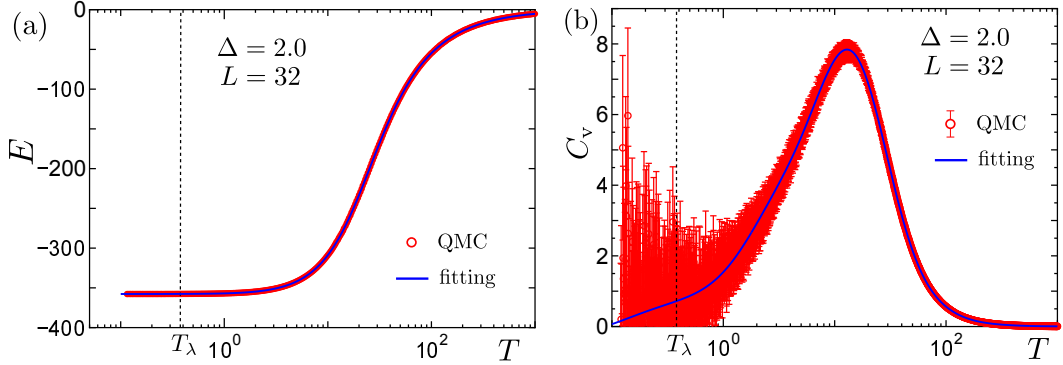}
\caption{
(Color online)
Gaussian kernel fittings of the internal energy $E$ and the specific heat $C_{\rm v}$ for $\Delta = 2.0$ and $L=32$.
(a)Fitting result for $E$, where error bars of the QMC data are smaller than the symbol size. 
Solid line represents the fitting result based on the Gaussian kernel method
(b)Fitting result for $C_{\rm v}$, where symbols with error bares represents the QMC results.
Solid line represents the fitting result based on the Gaussian kernel method.
The vertical dotted line indicates the lattice Unruh temperature $T_\lambda=0.3796\cdots$ for  $\Delta = 2.0$.
}\label{Cv_fit}
\end{figure}

In order to extract a reliable $S_{\rm EE}$ from the QMC results, we use nonlinear fitting  based on the kernel method.
Let us briefly summarize the Gaussian kernel method.
For a set of numerical data $(x_i, y_i, \delta y_i)$, where $i=1, \cdots d$, and $\delta y_i$ is the variance of $y_i$, we estimate a function $f(x)$ with the from 
\begin{align}
f(x) = \sum_{i, j} y_i \Sigma^{-1}_{ij} K(x, x_j),
\end{align}
where $K(x_i, x_j)$ and $\Sigma_{i,j}$ respectively represent a kernel function and a covariance matrix.
Here, we assume a Gaussian kernel
\begin{equation}
K(x_i, x_j) = h_0^2 \exp\left[-\frac{(x_i-x_j)^2}{2h_1^2} \right] 
 \label{eq_2}
\end{equation}
where $h_{0}$ and $h_1$ are hyper parameters.
We also assume the covariance matrix as
\begin{align}
\Sigma_{ij}= K(x_i, x_j) + \delta y_i^2 \delta_{ij}
\end{align}
with $\delta y_i$ being the variance of a Gaussian noise. 
We then determine the hyper parameters by minimizing the log-likelihood function defined by
\begin{equation}
{\cal L}=-\frac{1}{2}\log\left|2\pi\Sigma\right|-\frac{1}{2}\sum_{i, j}y_i\Sigma_{ij}y_j \label{eq_3} \,.
\end{equation}
Note that errors included in the hyper parameters determine the error of the estimated value of $f(x)$.

For the estimation of $S_{\rm EE}$, we set
\begin{align}
(x_i, y_i, \delta y_i) = (\log T_i, E_i, \delta E_i)\, \quad {\rm or} \quad (\log T_i, C_i, \delta C_i) 
\end{align}
with $i=1, \cdots , d$, where $d$ denotes the number of data points.
For the present case, we use $d\simeq 1000$ for $E$, which includes mirror data to stabilize the fitting, and $d \simeq 2000$ for $C_{\rm v}$ in the temperature range $ 0.12 \lesssim T  \lesssim 1.0\times 10^5 $.

In Fig. \ref{Cv_fit}, we show QMC results of $E$ and $C_{\rm v}$ with error bars for $\Delta = 2.0$ and $L=32$,
where we can see that the QMC data for $C_{\rm v}$ contain large statistical fluctuations.
In the figure, the results of the Gaussian kernel fitting for $E$ and $C_{\rm v}$ is also presented as solid curves.
The hyper parameters estimated are 
\begin{align}
h_0= 1.16\times 10^2 \pm 0.19\times 10^2, \quad h_1 = 1.08 \pm 0.10  \quad {\rm for} \quad E
\end{align}
and 
\begin{align}
h_0= 2.16 \pm 0.92 , \quad h_1 = 1.04 \pm 0.10  \quad {\rm for} \quad C_{\rm v}\, .
\end{align}
Note that the error for the fitted results in Fig. \ref{Cv_fit} are smaller than the width of the curve both for $E$ and $C_{\rm v}$ in the scale of the figures.
We further perform the numerical integration of Eqs. (\ref{CtoS1}) and  (\ref{CtoS2}), using the fitted curves.
Then, the result with the internal energy is
\begin{align}
S_{\rm EE} = 0.957 \pm 0.001
\label{S32value}
\end{align}
The result with $C_{\rm v}$ is $S_{\rm EE} = 0.93 \pm 0.01$, which is basically consistent with Eq. (\ref{S32value}).
However,  the accuracy by $E$ is much better than that by $C_{\rm v}$.
We thus adopt the value of Eq. (\ref{S32value}) as a reliable estimation of $S_{\rm EE}$ extracted from the QMC simulation for ${\cal K}$.
We finally summarized the size dependence of $S_{\rm EE}$ for $\Delta =2.0$ as Fig. 3 in the main text.
Also, $S_{\rm EE}$ for $\Delta =3.0$ is shown as Fig. \ref{figEE3} in this supplementary material.
We can confirm that the QMC results for $\Delta=3.0$  converge to the exact  bulk value within $L=32$. 

\begin{figure}[tb]
\includegraphics[width=7cm]{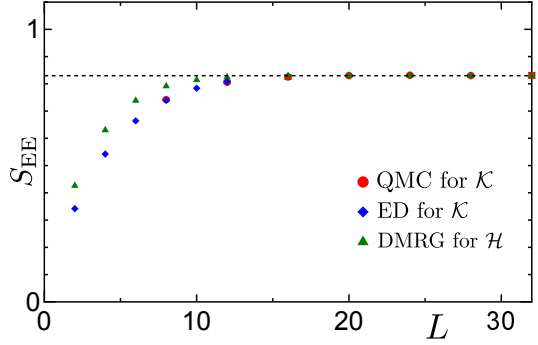} 
\caption{Entanglement entropy for $\Delta=3.0$ up to $L=32$.
Open circles indicate QMC results, where error bars are smaller than the symbol size. 
Exact diagonalization results for ${\cal K}$ up to $L=12$ are also shown as blue diamond symbols.
DMRG results of the entanglement entropy for the ground state of ${\cal H}$ are presented as green triangles for comparison.
The horizontal dotted line indicates the exact value $S_{\rm EE}=0.83025\cdots $.
}
\label{figEE3}
\end{figure}

\end{widetext}

\end{document}